\documentclass[12pt,preprint]{aastex}

\def\gtorder{\mathrel{\raise.3ex\hbox{$>$}\mkern-14mu
             \lower0.6ex\hbox{$\sim$}}}

\def\ltsima{$\; \buildrel < \over \sim \;$}
\def\simlt{\lower.5ex\hbox{\ltsima}}
\def\gtsima{$\; \buildrel > \over \sim \;$}
\def\simgt{\lower.5ex\hbox{\gtsima}}

\newcommand{\xray}{\mbox{X-ray}}
\newcommand{\hst}{\textit{HST}}

\begin{document}


\title{A high angular-resolution search for the progenitor of
       the type Ic Supernova 2004gt}


\author{Avishay Gal-Yam\altaffilmark{1}
D.~B.~Fox, S.~R.~Kulkarni, K.~Matthews, D.~C.~Leonard\altaffilmark{2}, 
D.~J.~Sand, D.-S.~Moon, S.~B.~Cenko, \& A.~M.~Soderberg}
\affil{Division of Physics, Mathematics and Astronomy, California Institute of 
     Technology, Pasadena, CA\,91125, USA}
\email{avishay@astro.caltech.edu}


\altaffiltext{1}{Hubble Fellow.}
\altaffiltext{2}{NSF Astronomy and Astrophysics Postdoctoral Fellow.}


\begin{abstract}

We report the results of a high-spatial-resolution search for
the progenitor of type Ic supernova SN\,2004gt, using the newly commissioned Keck
laser-guide star adaptive optics system (LGSAO) along with archival
\textit{Hubble Space Telescope} data.
This is the deepest search yet performed for the progenitor of any
type Ib/c event in a wide wavelength range stretching from the 
far UV to the near IR. We determine that the progenitor of SN\,2004gt
was most likely less luminous than $M_V=-5.5$ and $M_B=-6.5$ magnitudes.
The massive stars exploding as hydrogen-deficient
core-collapse supernovae (SNe) should have lost their outer hydrogen envelopes
prior to their explosion, either through winds -- such stars are identified 
within our Galaxy as as Wolf-Rayet (W-R) stars -- or to a binary companion. 
The luminosity limits we set rule out more than half of the known galactic
W-R stars as possible progenitors of this event. In particular, 
they imply that a W-R progenitor should have been among the more-evolved (highly
stripped, less luminous) of these stars, a concrete constraint on its
evolutionary state just prior to core collapse. The possibility of a less 
luminous, lower-mass binary progenitor cannot be constrained. This study
demonstrates the power of LGS observations in furthering our
understanding of core collapse, and the physics powering supernovae, GRBs and
XRFs.

\end{abstract}




\section{Introduction}

Stars with masses greater than about ten times the mass of the Sun end their
lives with a catastrophic core collapse that explodes the stellar
envelope, producing a core-collapse supernova (SN).  Understanding the
nature of these cosmic explosions in detail requires knowledge of the
properties of their progenitor stars.  The progenitors of the most
common type of core-collapse SNe (hydrogen-rich or type II) have
been identified (White and Malin 1987; Aldering, Humphreys, \& Richmond
1994; Van Dyk, Li, \& Filippenko 2003a; Smartt et al. 2003) as luminous supergiants.  

By contrast, the observational constraints on the progenitors of the
hydrogen-poor type Ib and Ic events (Filippenko 1997) are minimal.
Previous progenitor searches (Barth et al. 1996; Smartt et al. 2002; 
Van Dyk, Li, \& Filippenko 2003b; Maund \& Smartt 2005) utilizing pre-explosion 
images from the ground or from \textit{HST} have so far yielded upper limits that
were not sensitive enough to constrain the nature of the exploding star.   
This is unfortunate, because a small fraction of these events produce, 
as they die, not only a SN but also a gamma-ray
burst (GRB; e.g., Galama et al. 1998; Stanek et al. 2003; Hjorth et al. 2003; 
Malesani et al. 2004) or an \xray\ flash (XRF; Soderberg et al. 2005)
-- in the process, radiating an energy many times that of the
brightest SNe. Constraining the properties of evolved massive stars
that give rise to SNe Ib and Ic may lead to better understanding of the 
physics of these explosions, and may prove to be a key to successful 
modelling of GRBs and XRFs that are sometimes associated with such events.

Here, we present the deepest yet search for the progenitor of type Ic SN 2004gt,
using Keck laser-guide-star assisted adaptive-optics (LGSAO) observations ($\S~2$) to
pinpoint its location to within $0.06$ arc seconds and 
overcome source confusion in pre-explosion Hubble Space Telescope 
({\it HST}) data, a major limitation in some previous studies. We impose the first far UV, and deepest yet
optical and IR, upper limits on the luminosity of the SN progenitor ($\S~3$). 
These rule out a luminous ($M_V<-5.5$, $M_B<-6.5$)
Wolf-Rayet (W-R) star exploding as SN 2004gt. Similar results based on 
the localization of this SN using recent {\it HST} data, and an 
independent analysis of the archival data set, are presented by 
Maund et al. (2005) in this issue. 

\section{Observations}

On 28.48 January 2005 UT, forty seven days after the first
report (Monard 2004) about the type Ic (Ganeshalingam, Swift, 
Serduke, \& Filippenko 2004) supernova SN~2004gt, we undertook high angular resolution
observations in the 2.2-$\mu$m ($K_s$) band of the SN and its
vicinity using the wide-field channel (0.04 arcsecond/pixel)
of the Near Infra-Red Camera 2 (NIRC2) operated behind the
newly commissioned Laser Guide Star (LGS) assisted
Adaptive Optics (LGSAO; Wizinowich et al. 2004) system mounted on the Keck II
10-m telescope on Mauna Kea, Hawaii.  
An artificial beacon is created in the sky by 
shining a strong laser beam, tuned to the resonance line of sodium,
up into a sodium-rich layer, usually located at an altitude of
$\approx90$\,km. Using the beacon, the adverse blurring effects of the 
lower altitude atmosphere are measured and then corrected by a deformable 
mirror. The laser beacon is insensitive to tip-tilt corrections but 
we were able to use the SN itself for this purpose.
The data were reduced in the standard manner, with
bias-subtraction, flatfielding, fringe calculation and subtraction,
image registration, cosmic ray identification, cosmic ray and
bad-pixel masking, and final image combination performed using custom
software within the Pyraf environment.  
The resulting images, with full width at half maximum (FWHM) 
of 0.13\,arcseconds (more than 3 LGS pixels, so well-sampled), 
have exquisite resolution (Figure~\ref{fig:WIRC+LGS}), comparable 
to the angular resolution of images produced by the {\it Hubble 
Space Telescope} (HST).

HST observations of the vicinity of SN\,2004gt prior to
its explosion were extracted from the HST archive. Available data 
include Wide Field and Planetary Camera 2 (WFPC2) imaging in the 
F336W (4500s), F439W (4000s), F555W (4400s) and F814W 
(2000s) bands ($UBVI$ respectively, hereafter) 
obtained during January 20, 1996. The location of SN 2004gt falls on
the better-sampled (but less sensitive) PC chip, with pixel scale
0.05 arcsecond/pixel. Additionally, we examined a single 720s far-UV image 
obtained with the Space Telescope Imaging Spectrograph (STIS) using the far-UV MAMA
detector and the F25SRF2 filter, obtained on June 21, 1999. The
WFPC2 data were reduced by custom software using elements from the 
DRIZZLE (Fruchter \& Hook 2002) 
IRAF package (see Sand et al. 2005) and photometered using the
HSTPHOT package (Dolphin et al. 2000).  

\section{Upper limits on the luminosity of the progenitor of SN 2004gt}

Our LGSAO observations of SN~2004gt were motivated by three reasons.
First, the empirical constraints on progenitors of type Ic supernovae
are poor ($\S~1$). Second, there exists a treasure trove of
archival (pre-supernova) HST observations of NGC~4038/4039, covering
the entire UV/optical wavelength range (1500\,\AA\ to 9000\,\AA; $\S~2$). 
Third, the exquisite image quality of the LGSAO images means that we can
pinpoint the location of the SN, and hence its progenitor, among 
the numerous stars and star clusters visible in this part of the host galaxy
(see Figure~\ref{fig:WIRC+LGS}).

\begin{figure*}
\includegraphics[width=17cm]{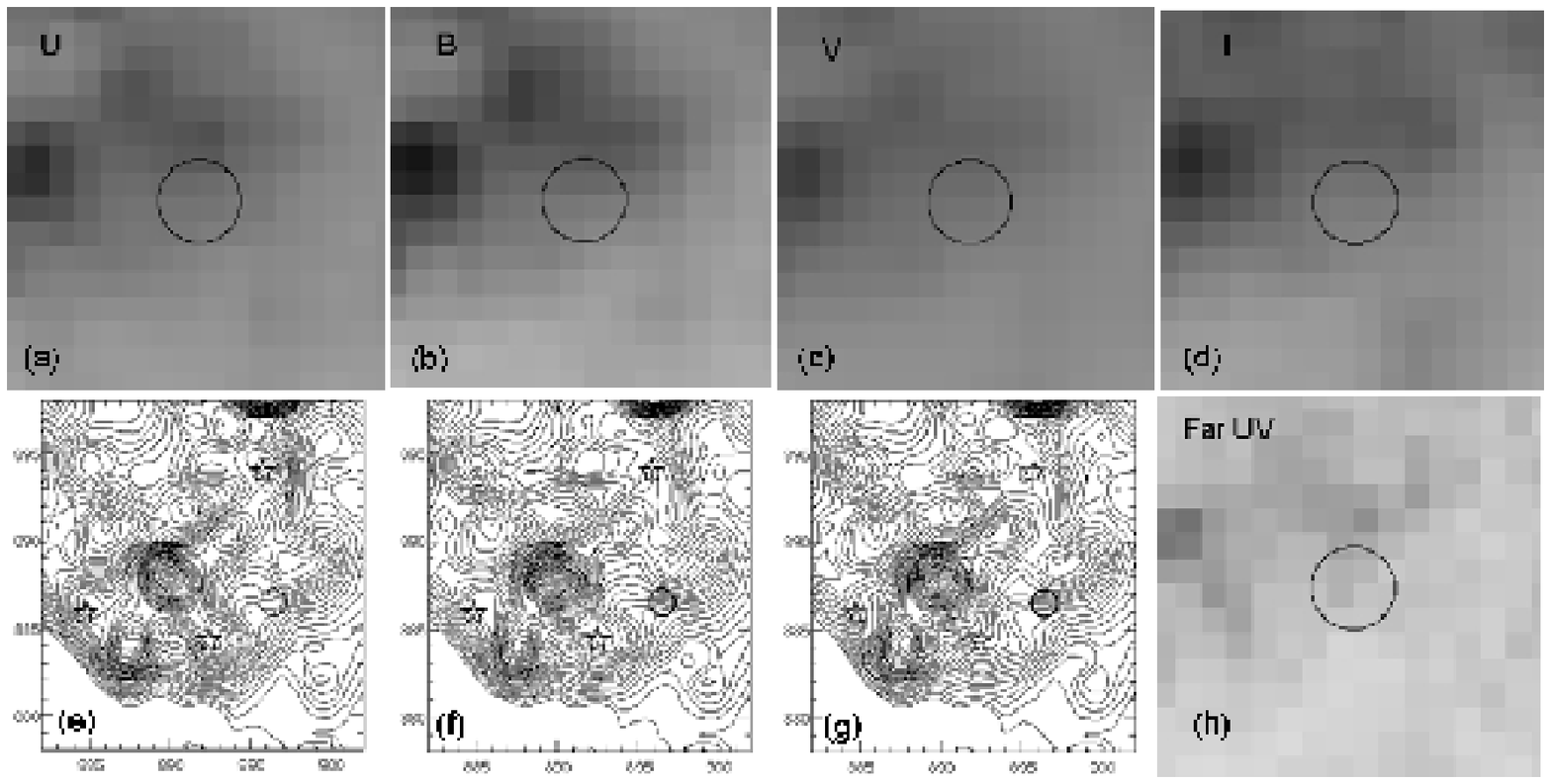}
\caption{
Pre-explosion \hst\ images of the immediate vicinity of SN\,2004gt, in
five bands, and illustration of our sensitivity estimation method.
Coadded \hst\ images in the $U$ (a), $B$ (b), $V$ (c), and $I$-bands
(d) shows no point-like source within the 5$\sigma$ localization
region for the SN. Panels (e)-(g) demonstrate our sensitivity estimation
(the $B$-band data are shown for example). 
We insert artificial sources with known flux into locations 
(stars in panels (e)-(g)) with background levels similar to those 
we measure at the position of SN 2004gt (circle). Panel (e) shows
a clear detection while panel (f) a marginal one. We set our limit as the 
brightest non-detection (g).    
We find that any putative progenitor had to be fainter
than apparent magnitudes, $m_U=22.7$, $m_B=24.65$, $m_V=25.5$, $m_I=24.4$ and
$m_{\lambda=150{\rm nm}}=23.4$.
Panel (a)-(d) and (h) are $0.64 \times 0.64$ arcseconds across, while panels
(e)-(g) are $0.8 \times 0.8$ arcseconds. In all cases North is up and East 
is to the left.
} 
\label{fig:HST}
\end{figure*}

We register the Keck-LGS $K_s$-band and archival \hst\ $I$-band
images in the following manner. 
We choose the $I$-band \hst\ image to minimize the effect
of any wavelength-dependent centroid shifts for objects used in the
registration.  We perform coarse alignment using six bright sources,
and then make a finer alignment using the eleven sources indicated by
circles in panels Figure~\ref{fig:register} (a) and (b).  
The IRAF tasks ``geomap'' and
``geotran'' are used throughout, and a general second-order 
transformation is applied.  
We then register the remaining \hst\ images against the $I$-band image
using the same tasks.  We determine the
final uncertainty in the SN location on the pre-explosion \hst\ images
by adding in quadrature the SN centroiding error
(Fig.~\ref{fig:WIRC+LGS}), the RMS error in the \hst\ $I$-band to Keck
$K_s$-band registration, and (where relevant) the RMS error in
registration between the \hst\ $I$-band image and the target image.
We find 1$\sigma$ total uncertainties of 0.24 to 0.31 Keck (LGS)
pixels depending on the band ($\sim$0.01\,arcsecond).
Figs.~\ref{fig:WIRC+LGS}(b) and \ref{fig:HST} show the final 5$\sigma$
error circles.

As can be seen from
Figure~\ref{fig:HST}(a)-(d) we could not identify any distinct progenitor
star at the location of SN~2004gt in any of the available archival HST 
images. Following Gal-Yam et al. (2004a) we estimate our limiting magnitude for an
associated point source (progenitor star) as follows.  We derive our
photometric zero-points from bright and well-isolated stars using the
\textit{HSTPHOT} package (Dolphin 2000). Introducing
artificial stars of progressively fainter magnitude (panels e-g, where
the $B$-band image is used for illustration) at locations (indicated
by stars) with background levels consistent with the location of
SN\,2004gt (indicated by the circle), we determine the limiting
magnitude for detection of a point-source (f), as contrasted with the
brightest non-detection (g).  We derive limiting magnitudes of
$m_U=22.7 \pm 0.5$, $m_B=24.65 \pm 0.15$, $m_V=25.5 \pm 0.15$ and
$m_I=24.4 \pm 0.3$, where our errors are dominated by the uncertainty
in the local zero point.
The far-UV STIS data (h) present a special case, as this
detector-filter combination has not been extensively calibrated.
First, we note that although a single pixel near the SN location
appears elevated, it is only a 1.5$\sigma$ excursion above the
brightness of nearby pixels and is derived from a single image.
Second, our estimated limiting magnitude of $m_{far UV}=23.4$ is
calculated via photometric coefficients in the image header ($PHOTFLAM
= 3.951175 \times 10^{17}$ and $PHOTZPT = -21.1$), and using a
0.5\,arcsecond aperture.

Adopting the most recently determined distance to the Antennae
galaxies, $d=13.8 \pm 1.7$ Mpc (Saviane, Hibbard, \& Rich 2004) 
the magnitude limits result in the progenitor star being fainter 
than $1.25 \times 10^4$ solar luminosities in the $V$ band.  
The corresponding absolute magnitudes are $M_U> -8.0$, $M_B >-6.05$, 
$M_V >-5.2$ and $M_I >-6.3$.  These limits are $\approx 2-10$ times 
tighter than any previously obtained ($\S~1$). 

It is important to quantify the effect of interstellar dust between us and the
supernova. In this direction interstellar extinction
by Galactic dust is small, $E_{B-V}=0.012$ mag (Burstein \& Heiles 1982) 
to 0.046 mag (Schlegel, Finkbeiner, \& Davis 1998). Whitmore
et al. (1999) report their photometric and spectroscopic analysis is 
consistent with the lower value of the two which they consider negligible, 
and which we also adopt (but do not neglect) here. In the same work, the
extinction, inferred from HST ultra-violet (UV) spectroscopy, of
the star cluster clearly seen in the vicinity of SN~2004gt
(Figure~\ref{fig:WIRC+LGS}) is also low, $E_{B-V}=0.01 \pm 0.04$ mag
(Whitmore et al. 1999).
The equivalent width, $W$, of the Na~I D-absorption lines in SN
spectra has been argued (Turatto, Benetti, \& Cappellaro 2002) 
to be a tracer of interstellar
gas (and thus also of dust). From a spectrum of SN~2004gt 
obtained using the DBSP spectrograph mounted on the 200" Hale
telescope at Palomar Observatory as part of the CCCP program 
(Gal-Yam et al. 2004b) we measure $W=0.6\,$\AA,
leading to $E_{B-V}=0.09$ to 0.27\,mag.  
In view of the small extinction towards the nearby
cluster, we favor the lower value of extinction for the SN
itself. Thus, assuming $E_{B-V}=0.012$ mag and $E_{B-V}=0.09$ mag 
for the Galactic and host extinction, respectively,
we find the extinction-corrected limits $M_U> -8.6$, $M_B >-6.5$, 
$M_V >-5.5$ and $M_I>-6.5$. 

\section{Discussion and conclusions}

SNe of type Ib and Ic exhibit no hydrogen in their spectrum.
Their progenitors must be massive stars which have lost their
hydrogen envelope. This could result from wind-driven mass-loss 
is very massive stars, $M>25-40 M_{\odot}$ according to different models (see,
e.g., Maeder \& Conti 1994 for a review) -- such stars are identified locally as
Wolf-Rayet (W-R) stars. Alternatively, a lower-mass star in a close binary system
may lose its outer envelope through interaction with its companion 
(see Podsiadlowski et al. 2004 and references therein).
 
Within the W-R class, the observed range of spectral properties has been
further mapped onto an evolutionary path (Maeder \& Conti 1994).  
Of those W-R stars whose spectra are dominated by helium and nitrogen features,
consistent with products of hydrogen burning via the Carbon-Nitrogen-Oxygen
cycle, the cooler but more luminous (``late'') WNL stars are believed
to be less evolved than the hotter, but less luminous (``early'') WNE stars.
Members of the WC and WO subclasses, whose carbon and oxygen-dominated
compositions are indicative of helium fusion products, are thought to be more
evolved still, displaying material from the deeper stellar layers on their 
highly stripped surface.

Our observations allow us to conclude that the progenitor of SN~2004gt
was fainter than the median star in the compilation (Vacca \& Torres-Dodgen
1990) of Galactic WR stars\footnotemark. 
\footnotetext{We correct the narrow-band $v$ magnitudes given by Vacca \& 
Torres-Dodgen (1990) to broad-band (Johnson) $V$ magnitudes (similar to the
F555W WFPC2 filter we analyze) by applying individual $V-v$ corrections
to each star which we derive using the SIMBAD database.}  
More importantly, on luminosity grounds, the
progenitor is not a WNL star, since, adopting our estimated
extinction and distance to the Antennae, we would have detected 
every WNL star listed by Vacca and Torres-Dodgen (1990).  This clue -- that the progenitor
of SN 2004gt, if a W-R star, must have been a WNE, WC or WO star -- directly suggests that
type Ic supernova explosions, deficient in both hydrogen and
helium, result from evolved progenitors. An immediate corollary would
be that progenitors of type Ib SNe (hydrogen poor, but helium-rich)
could be among the less-evolved and more luminous WNL stars. 
This finding is in general accord
with our current theoretical understanding of how stars evolve, and
constitutes the first direct support of this picture. 
We note that SN Ib/c models invoking less massive, lower-luminosity
progenitors in binary systems, cannot be constrained by our observations.

The observations presented here bode well for making rapid progress
in relating the types of SNe to their progenitors, a subject
currently dominated primarily by theory and models. LGS observations
on large-aperture telescopes have the necessary sensitivity and 
precision to routinely localize SNe relative
to archival high resolution images of their host galaxies. Currently
the {\it Hubble Space Telescope} provides the best archive of nearby
galaxies. However, as LGS methodology becomes more common it is
quite conceivable to undertake a dedicated program of imaging the $\approx1000$ 
nearest galaxies known to contain significant populations of massive stars,
in preparation for future SNe. The goal of this effort would be to
empirically link the nature of exploding stars and the properties of
the resulting explosions, from which we can understand the
physical processes that govern one of Nature's most
dramatic events, the explosions of dying stars as SNe,
gamma-ray bursts and X-ray flashes.

%

\section*{Acknowledgments}

A.G. acknowledges support by NASA through Hubble Fellowship grant
\#HST-HF-01158.01-A awarded by STScI, which is operated by AURA, Inc.,
for NASA, under contract NAS 5-26555.  D.C.L. is supported by a
National Science Foundation (NSF) Astronomy and Astrophysics
Postdoctoral Fellowship under award AST-0401479.
A.M.S. and D.J.S are supported through the NASA Graduate Student 
Research Program, under NASA grant NAGT-50449.
SRK's research is supported by NSF and NASA.
We thank C. Gelino, D. Maoz and G. P. Smith for help and advice.
We acknowledge helpful comments from an anonymous referee.

\clearpage

\begin{figure*}
\caption{
\newpage
{\bf (a)} Composite near-infrared ($JHK_s$) image of SN\,2004gt and
its host galaxy, the Antennae, taken on 16 January 2005 UT within the
context of our ongoing CCCP program at Palomar
Observatory (Gal-Yam et al. 2004b). The SN (centered within the red square) is
barely resolved from the bright cluster ``S'' to its
northwest (Whitmore et al. 1999).
{\bf (b)} The vicinity of SN\,2004gt (red square in panel a), prior to
its explosion, as seen in the far-UV to $I$-band by {\it HST}.  
The color composite
uses far UV (STIS 1450~\AA) and $U$-band (WFPC2 $3360$~\AA) images for
blue, $B$-band (WFPC2 $4390$~\AA) and $V$-band (WFPC2 $5550$~\AA)
images for green, and the $I$-band (WFPC2 $8140$~\AA) image for red.
The yellow circle gives our 5$\sigma$ uncertainty in the SN position
on this image (see Fig.~\ref{fig:register}).
{\bf (c)} SN\,2004gt and its immediate vicinity at high-resolution in
$K_s$-band (0.13\,arcsecond full-width at half-maximum). The SN location 
(indicated by the cross) is determined to $<$0.05\,pixel precision
(0.002\,arcsecond) using the centroid finding algorithm within IRAF.
} 
\label{fig:WIRC+LGS}
\end{figure*} 

\clearpage

\begin{figure*}
\includegraphics[width=17cm]{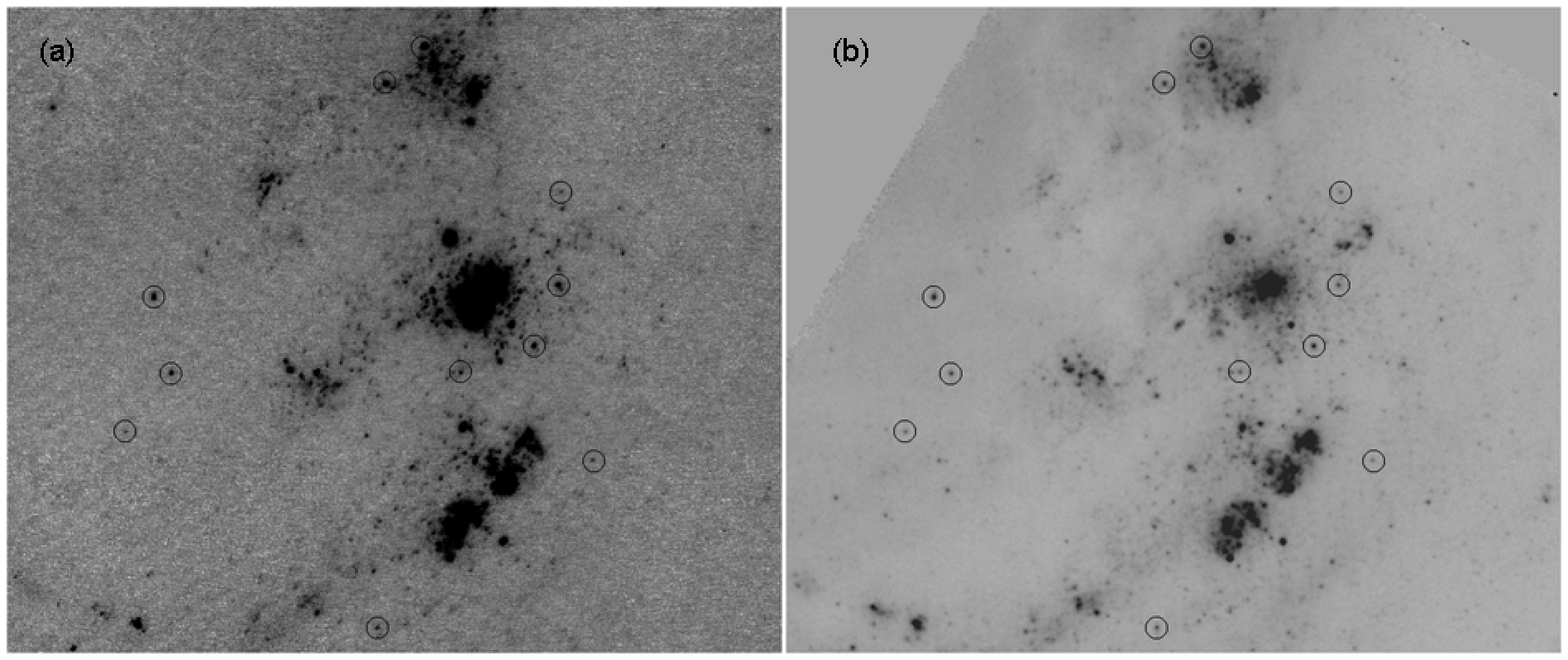}
\caption{
Registration of Keck-LGS $K_s$-band (a) and archival \hst\ $I$-band
(b) images. Using 11 stars common to archival HST images and our LGS image (circled) we are
able to register SN~2004gt onto pre-explosion images, 
to within $<$0.06\,arcsecond, at 5-$\sigma$
confidence level. 
Figs.~\ref{fig:WIRC+LGS}(b) and \ref{fig:HST} show the final 5$\sigma$
error circles. Each panel is $28 \times 24$ arcseconds, North is up and East to 
the left.
} 
\label{fig:register}
\end{figure*}

\clearpage

\end{document}